\documentclass[fleqn,usenatbib]{mnras}

\usepackage{newtxtext,newtxmath}

\usepackage[T1]{fontenc}

\DeclareRobustCommand{\VAN}[3]{#2}
\let\VANthebibliography\thebibliography
\def\thebibliography{\DeclareRobustCommand{\VAN}[3]{##3}\VANthebibliography}

\usepackage{graphicx}	
\usepackage{amsmath}	

\usepackage{subfig}

\title[Stellar prominences and mass loss rates]{Influence of magnetic cycles on stellar prominences and their mass loss rates}

\author[S. J. Faller et al.]{
Sarah J. Faller,$^{1}$\thanks{E-mail: sjf8@st-andrews.ac.uk (SJF)}
Moira. M. Jardine,$^{1}$
\\
$^{1}$School of Physics and Astronomy, University of St Andrews, North Haugh, St Andrews, Fife, Scotland, KY16 YSS\\
}

\date{Accepted XXX. Received YYY; in original form ZZZ}

\pubyear{2022}

\begin{document}
\label{firstpage}
\pagerange{\pageref{firstpage}--\pageref{lastpage}}
\maketitle

\begin{abstract}

Observations of rapidly-rotating cool stars often show coronal “slingshot” prominences that remove mass and angular momentum when they are ejected. The derived masses of these prominences show a scatter of some two orders of magnitude. In order to investigate if this scatter could be intrinsic, we use a full magnetic cycle of solar magnetograms to model the coronal structure and prominence distribution in a young Sun, where we scale the field strength in the magnetograms with angular velocity according to $B \propto \Omega^{-1.32}$. We reproduce both the observed prominence masses and their scatter. We show that both the field strength and the field geometry contribute to the prominence masses that can be supported and to the rate at which they are ejected. Predicted prominence masses follow the magnetic cycle, but with half the period, peaking both at cycle maximum and at cycle minimum. We show that mass loss rates in prominences are less than those predicted for the stellar wind. We also investigate the role of small-scale field that may be unresolved in typical stellar magnetograms. This provides only a small reduction in the predicted total prominence mass, principally by reducing the number of large magnetic loops that can support slingshot prominences. We conclude that the observed scatter in prominence masses can be explained by underlying magnetic cycles.

\end{abstract}




\section{Introduction}

Prominences are structures that form in the corona of a star. They are condensations of relatively cool gas that are supported in the coronal magnetic field and their locations and dynamics provide a wealth of information about the coronal environments in which they are embedded. On the Sun, they are observed as long, thin filaments that appear dark against the solar disk as they scatter surface photons out of the line of sight of an observer. Solar prominences are seen at maximum heights of 0.5$R_{\odot}$  above the surface and quiescent solar prominences have mean heights of $2.6 \times 10^4 $km \citep{2010ApJ...717..973W}, and masses on the order of $10^{12} - 10^{15}$g \citep{Labrosse2010}.

When the Sun was younger and rotated more rapidly, however, it may also have supported larger ``slingshot'' prominences  \citep{2019MNRAS.485.1448V}. These clouds of cool gas are observed as transient absorption features in  H$\alpha$ that recur on successive rotations of the star \citep{10.1093/mnras/238.2.657}. This indicates that the absorbing material is co-rotating with the star. The radial acceleration of these features is a direct measure of the distance of the absorbing mass from the stellar rotation axis.  These prominences are typically found to be trapped around, or beyond, the co-rotation radius $R_k = (GM_\star/\Omega^2)^{1/3}$ which may be several stellar radii from the stellar rotation axis. The absorbing clouds must therefore be confined against centrifugal ejection, presumably by closed magnetic loops. 

These ``slingshot'' prominences were first observed on AB Doradus \citep{10.1093/mnras/238.2.657} but have since been detected on many other rapidly-rotating stars such as Speedy Mic \citep{10.1093/mnras/262.2.369,10.1111/j.1365-2966.2006.11128.x},  HK Aqr \citep{Byrne1996,10.1093/mnras/stw1922} and PZ Tel \citep{Barnes2000, 10.1093/mnras/stw1922}, and AP 149 \citep{cang_petit_folsom_donati_2019}. The fact that these prominences are frequently observed on rapid rotators suggests that the  closed field lines that support them must also extend out to large distances from the stellar surface. \citet{2019MNRAS.483.1513W} show that support of slingshot prominences can be achieved within relatively simple field configurations - a necessity as the lowest order field modes are dominant out at at co-rotation. 
The prominence-bearing loops may lie within the X-ray emitting corona
(whose extent can be revealed by the velocity shifts of X-ray lines \citep{2007MNRAS.377.1488H}) or  be formed by reconnection of the stellar wind field lines above the summits of helmet streamers \citep{2005MNRAS.361.1173J}. 

While prominences are useful tracers of coronal structure, this can also be revealed by interferometric observations. Recently, VLBI observations of AB Dor have revealed lobes of emission extending out between 5 and 18 stellar radii from the star \citep{2020A&A...641A..90C}. It is thought the source of these compact radio emissions are slingshot prominences. The authors discuss possible configurations of prominences which could lead to the observed radio emission, including the possibility of support above a helmet streamer. Such support mechanisms are discussed by \citet{2005MNRAS.361.1173J}, who postulate that prominences may be supported in the stellar wind. A similar, but more extended, configuration was observed by \citet{2008A&A...480..489M} between the T Tauri binary stars comprising V773 Tau A. Each star appeared to support structures out to 18 $R_{\star}$ in parallel helmet streamers. These helmet streamers interact to induce flaring due to the compression of the streamers' magnetic fields. It is possible that these helmet streamers each support prominences.   Earlier VLBI observations of the dMe star UV Ceti also showed two lobes of emission which the authors attributed to regions above the rotation poles of the star \citep{10.1046/j.1365-8711.2002.04972.x}.

Stellar slingshot prominences also differ from solar prominences in their masses. These can be found from the examination of the H$\alpha$ and Ca H\&K lines. Prominences are optically thick in H$\alpha$  so the depth of the transit profile is used to calculate the area of the prominence. The masses of prominences are then a product of the prominence area, the Hydrogen column density and the atomic mass of Hydrogen \citep{10.1111/j.1365-2966.2006.11128.x}. Typical masses for AB Dor prominences are $2-6\times 10^{17}$g \citep{1990MNRAS.247..415C} and similar values are found for individual prominences on other stars. For example, Speedy Mic prominences have masses of $0.5-2.3 \times 10^{17}$g \citep{10.1111/j.1365-2966.2006.11128.x}. Prominences are also observed on rapidly-rotating M dwarfs - \citet{10.1093/mnras/stw1922} give the mass of a single prominence on HK Aqr as $5.7 \times 10^{16}$g while for V374 Peg masses greater than $10^{16}$g have been found \citep{2016A&A...590A..11V}. More recently, YZ CMi showed a prominence eruption ejecting material with mass in the range $10^{16}-10^{18}$g \citep{2021PASJ...73...44M}.

Rapidly-moving features such as these also reveal some of the dynamics of stellar coronae. 
\citet{1999A&A...341..527E} provide an analysis of H$\alpha$ chromospheric emission of the star BD+$22^{\circ}4409$, which shows a broadened profile consistent with continuous downwards flow of absorbing material towards the pole of the star. The authors suggest that coronal condensations are the source of these downflows. This creates a system of transferring material between the chromosphere and the corona. Material that condenses below the co-rotation radius, but which does not find a stable location in which to accumulate into  a prominence, simply falls back to the stellar surface.  BD+$22^{\circ}4409$ is seen pole-on so if the star is able to host slingshot prominences then they would not be seen to transit the star.

In addition to red-shifted H$\alpha$ transients, there are also some that appear highly blue-shifted such as those observed on YZ CMi \citep{2021PASJ...73...44M}. 
These may be indicators of material that is ejected from a star \citep{Houdebine1990,GuentherEmerson1997,2016A&A...590A..11V,FuhrmeisterSchmitt2018, Namekata2021}. A survey of a sample of 25 rapidly rotating M-dwarfs showed numerous absorption transients, both blue- and red-shifted \citep{2019A&A...623A..49V}.
The blue-shifted absorption transients were thought to be CMEs but their velocities were below what we observe for solar CMEs. The source of these features remains a puzzle, but their existence points to dynamic processes in the stars' coronae.

Material ejected from a star carries away not only mass, but angular momentum. Any prominences that form above the co-rotation radius will be ejected when they are destabilised and so will contribute to the overall spin-down of the star. The majority of angular momentum loss is believed to be due to the wind, yet between different wind models there is a significant variation in predicted torques. 

\citet{articlereville} performed a study on 5 K-type stars and the Sun and show that the predicted torque in the wind is insufficient to explain spin-down timescales.
Additional sources of mass and angular momentum loss outside of the wind are seemingly essential to explain the discrepancy. This additional angular momentum loss may be via CMEs or prominence ejection. Solar mass loss from CMEs is an order of magnitudes less than the background solar wind, but on other stars the relative contribution from CMEs may be much greater.
\citet{2017ApJ...840..114C} suggests that before the Sun reached the age of 1Gy the CME mass loss rate could have been 10-100 times greater that the wind mass loss; an important consideration for young solar-type stars.

\citet{articleaarnio} developed an empirical relationship to compare CME mass with X-ray flare energy in pre-main sequence solar-type stars. They calibrated the relationship using solar data, comparing observed CME mass-loss with the values predicted using X-ray emission. The authors quote a lower limit of current-day solar CME mass-loss to be $7.8 \times 10^{-16} $ M$_{\odot}$ yr$^{-1}$, yet calculate CME mass-losses in the range $10^{-12}-10^{-9} $ M$_{\odot} $ yr$^{-1}$ for solar-type pre-main sequence stars. They conclude that CME mass-loss rates above the critical value of $10^{-10} $ M$_{\odot}$ yr$^{-1}$ during the T Tauri phase is sufficient to influence the rotational evolution of a star.

\citet{2018ApJ...864..125F} compare MHD based models to surface magnetogram models and find nearly an order of magnitude difference in the average solar wind torques. This discrepancy is attributed to the difficulty in determining the amount of open flux from magnetograms, which is an important factor in wind calculations. The authors compare a third method based on the predicted spin down of solar type stars and find an even larger predicted value. 

Wind torques are functions of mass loss, which is very difficult to measure and observation attempts often result in a non-detection. \citet{2019MNRAS.482.2853J} propose using slingshot prominences as ``wind-gauges''. Prominences are formed when plasma from the stellar surface flows along closed field lines and is caught in stable points in the coronal magnetic field. These up-flows are similar in nature to the wind. If the mass and lifetime of a prominence is known then the rate at which the prominence fills provides an estimate of the wind mass flow rate. However, this method requires an assumption of the fraction of the stellar surface area that feeds the prominence.

\citet{2021ApJ...915...37W} gives a comprehensive summary of successful measurements of stellar wind mass loss rates made using the ``astrosphere method''. These show a positive power-law relation between mass loss rate per unit surface area with stellar X-ray flux. These observations provide strong constraints on the structure of the coronae and winds of these stars, which have been formalised through a series of power law relations between the mass loss rates and the fundamental parameters (the magnetic field strength, base density and temperature) \citep{2020A&A...635A.170A}. The authors use the relation of mass loss to X-ray flux with the dependence of X-ray luminosity on rotation to link stellar magnetic field strength and mass loss rates to Rossby number and stellar mass.

The observed wind mass loss rates presented in \citet{2021ApJ...915...37W} do show significant scatter, however, both between stars of different masses or rotation rates and for multiple observations of the same star. For many stars there is only one value obtained. From {\it in-situ} measurements we know that the solar wind varies in time due to the nature of the magnetic cycle. Changes to the surface magnetic flux and coronal field structure are intrinsic causes of variation in the wind mass loss rates. 
Also, the solar wind is not uniform across the entire solar surface. Coronal holes such as polar regions are sources of fast solar wind whereas latitudes near the equator are sources of slow wind. Stellar observations are limited to one viewing inclination, so it is unclear if their winds are also spatially variable. The observed scatter may therefore be intrinsic, or a result of the nature of the observational method.

Scatter in observations and predictions is also abundant for prominence masses.
\citet{2018MNRAS.475L..25V} use a simple model of prominence support to show that the prominence mass $m_p$ can be expressed as  $m_p/M_\star = (E_M/E_G)(r_\star/R_K)^2F$ where $E_M = B_\star^2R_\star^3$ and $E_G = GM_\star^2/R_\star$ are magnetic and gravitational energies and F is a geometric factor. Using published stellar parameters of stars in the mass range 0.1-1.34 $M_\odot$ that are predicted to host prominences, the authors predict a range of masses for individual prominences spanning four orders of magnitude. Both the stellar rotation rate and the magnetic field strength will evolve with stellar age.  Using rotational evolution models to predict prominence masses at a range of ages, \citet{2019MNRAS.485.1448V} predict a variation in the masses of individual prominences that is consistent with observations of individual stars.

These observations are of course snapshots in time that may not capture any variation due to short-term evolution of the magnetic field.
\citet{2020MNRAS.491.4076J} use surface magnetograms for AB Doradus obtained over several years to model the coronal field structure and to predict both the total mass of prominences supported and the fraction that is visible. The total predicted prominence mass varied by 2 orders of magnitude. Over the time span covered by the observations, the latitude of the dipole magnetic axis of the star also varied, changing not only the mass that can be supported, but also the fraction that is visible. The authors show that the inclination of the dipole axis alone can account for a large variation in total prominence mass. 
 
One limitation of using surface magnetograms to model prominence support is that they capture only the large-scale field \citep{2018MNRAS.478.4390L,2019MNRAS.483.5246L}. The small scale, more complex, field is unresolved and its role in influencing prominence support is as yet unknown. The role of small-scale field has been investigated in the context of stellar winds \citep{2015ApJ...807L...6G,2015ApJ...813...40G,2017MNRAS.465L..25J}. 
\citet{VSeeDipoleTorque} show that the dipole field is responsible for the majority of wind spin down torque, if the mass-loss rate is below some critical value. Above this value, however, smaller scale (higher order) fields make significant contributions. 
 
In this work, we aim to determine the extent to which the contribution of small-scale field or solar-like magnetic cycles influences the loss of mass and angular momentum from rapidly-rotating stars. We use solar magnetograms obtained throughout a solar cycle as inputs to a model for the varying coronal structure of a young, rapidly-rotating solar-like star. Based on this, we determine the cyclic variation in the prominence support and the mass and angular momentum loss that may come from prominence ejection or from the wind.

\section{Methods}
\label{sec:methods} 
\begin{figure*}
    \centering
    
        \includegraphics[width=8.0cm]{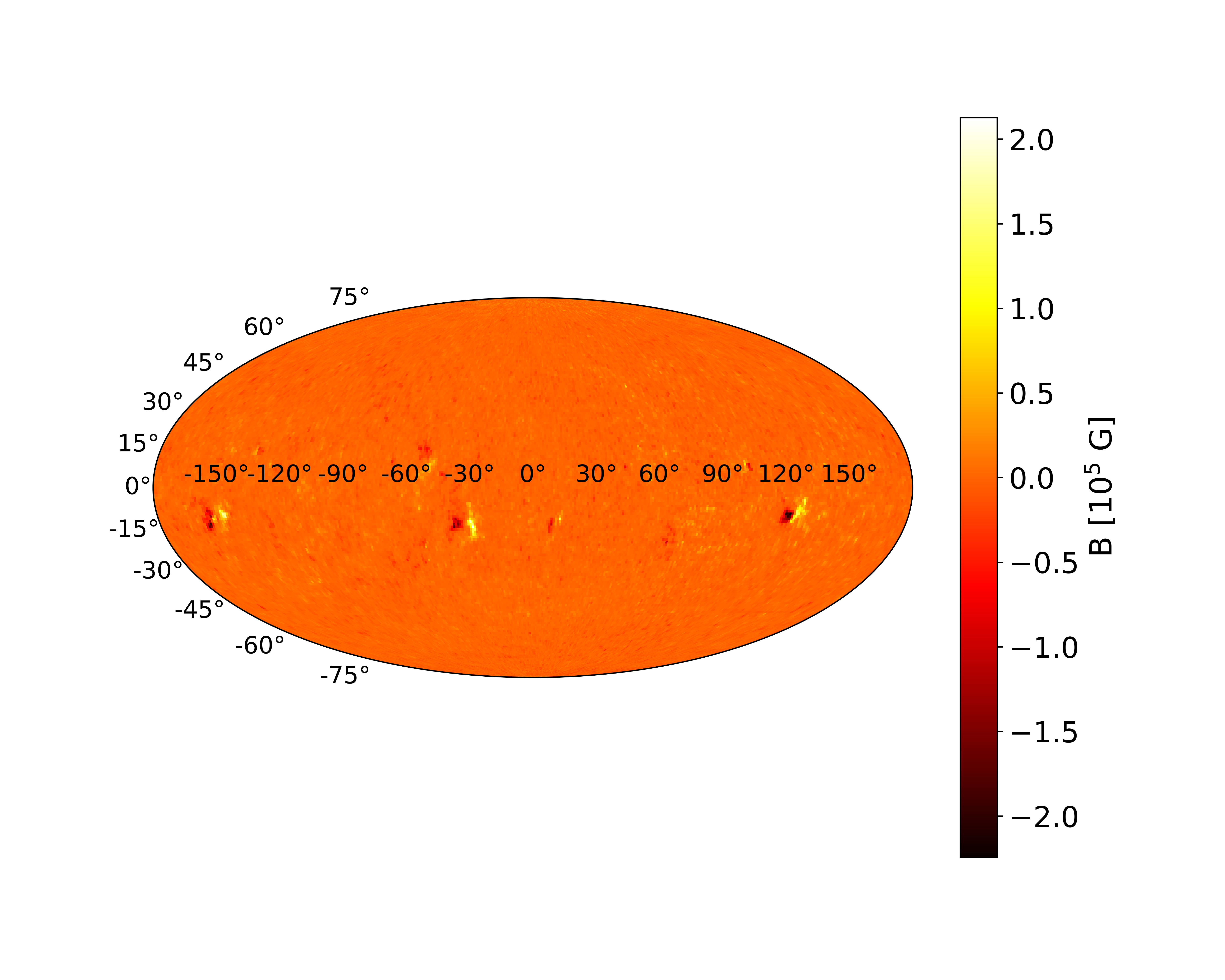}
        \includegraphics[width=8.0cm]{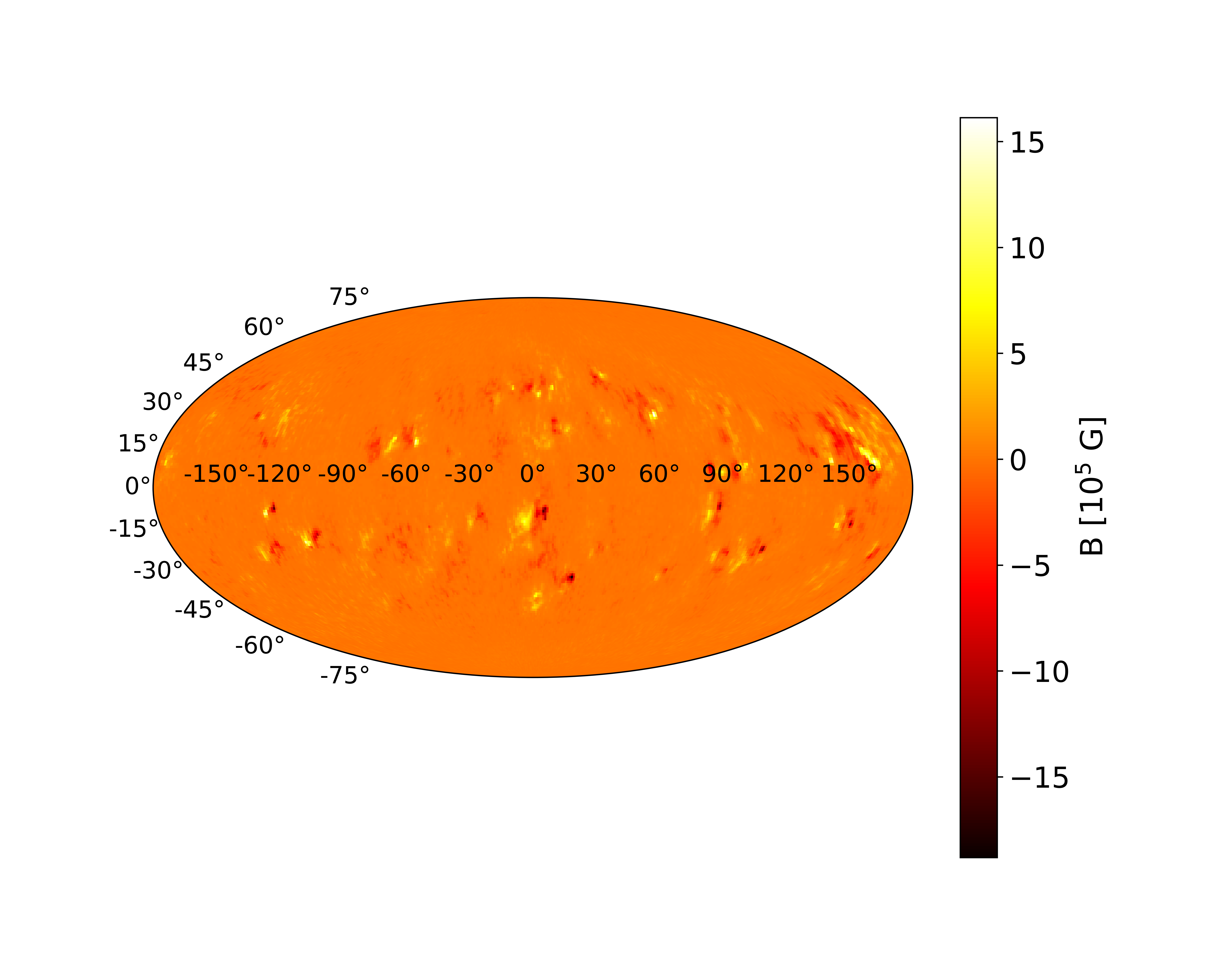}

        \includegraphics[width=8.0cm]{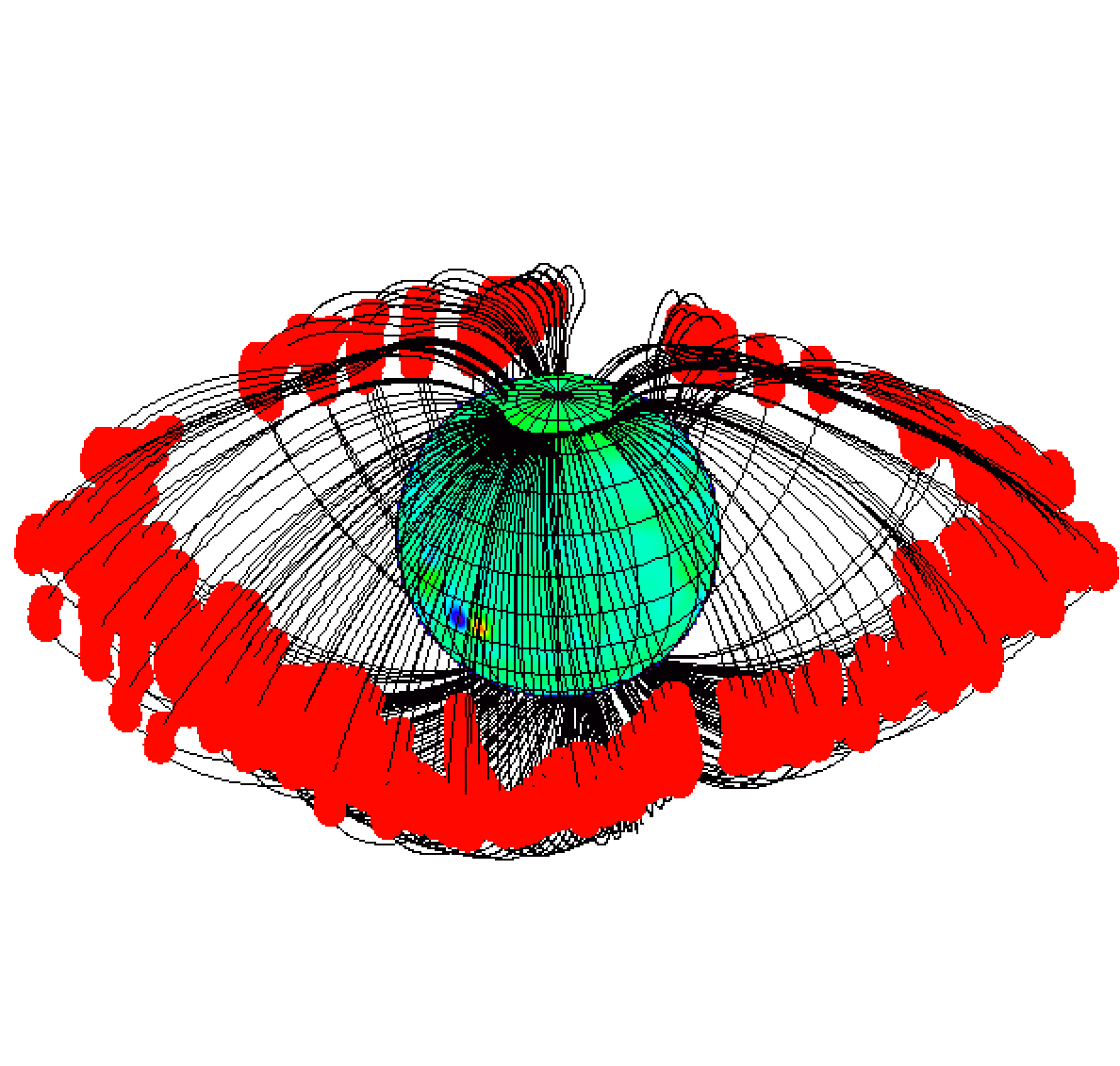}
        \includegraphics[width=8.0cm]{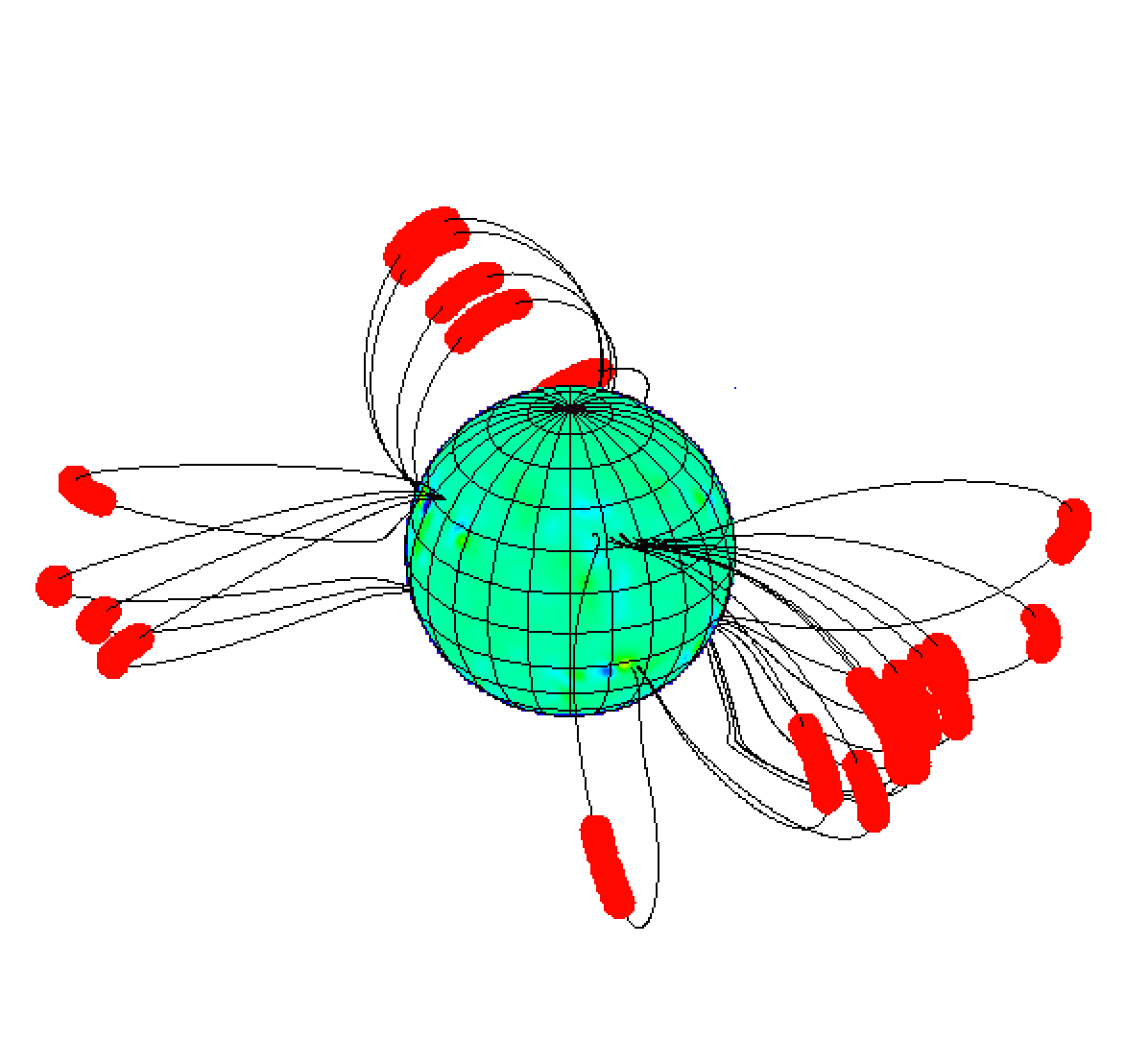}
    \caption{The top row shows magnetograms close to cycle minimum CR1628 (left) and cycle maximum CR1703 (right), and the colour bar represents the surface magnetic field in units of $10^5$ Gauss. The bottom  row shows the resulting prominence distribution in red, and the field lines that support prominences in black.
}
    \label{fig:prom_models}
\end{figure*} 
\subsection{Data}

Magnetograms from the US National Solar Observatory, Kitt Peak, were used to construct a series of coronal magnetic fields for a theoretical young solar-type star, based on the mass and radius of AB Doradus.  The maps cover one solar magnetic cycle, from February 1975 - April 2000. Each magnetogram gives the radial line of sight component of the magnetic field at the solar surface. The background variation was removed to isolate only the cyclic behaviour.

To simulate a young star with a short rotation period, the surface field strength was uniformly increased using the relation $B \propto \Omega^{-1.32}$ where the rotation period was set to $0.515$ days, appropriate for AB Doradus \citep{2014MNRAS.441.2361V}. Along with the increased magnetic field strength, the coronal temperature was set to that of AB Doradus $ 8.57 \times 10^6 $ K \citep{2015A&A...578A.129J}. The maximum extent of the closed corona was set to 3.5 R$_{\star}$, just beyond the corotation radius (where most prominences are observed).  \par

The magnetograms give the radial component of the surface magnetic field that can be expressed as a summation series of spherical harmonics. This series may be truncated at any point to artificially remove smaller-scale (higher-order) field from the magnetograms. The spatial scale of the magnetic field depends on the maximum spherical harmonic, $l_{max}$ of the field decomposition. In degrees, the minimum spatial scale at the stellar surface is $180^{\circ}/l_{max}$.
In this work, 3 resolutions are considered: a simple dipole, a typical stellar resolution and a much higher resolution, which cannot at present be recovered for stars. At dipole resolution, only the largest scale field is recovered, so that we may see the inclination of the dipole axis relative to the stellar rotation axis, but no features such as active regions. At stellar resolution, it is expected that large spots and active regions will be resolved, but no smaller features. At solar resolution, we are able to see small sunspots and the small network of features called plage that surround sunspots. We use the notations LM01, LM08, and LM63 to denote spatial resolutions corresponding to $l_{max} = 1, 8,$ and $63$, and spatial scales of $\sim180, 23,$ and $3^{\circ}$ respectively.

\subsection{Coronal Field Structure}
The Potential Field Source Surface (PFSS) method \citep{AltschulerNewkirk1969} may be used to convert the surface values into a fully 3D description of the coronal magnetic field. The field is assumed to be current-free, and therefore potential.
In the current-free approximation, Ampere's law reduces to $\nabla \times \underline{B} = 0$, and the field can be expressed as the gradient of a scalar potential, $ \underline{B} = -\nabla\Psi$. 
With Gauss' law,  $\nabla\cdot\underline{B} = 0$, the potential,$\Psi$, is a solution to Laplace's equation, $\nabla^2\Psi = 0 $.

The potential in spherical co-ordinates is
\begin{equation}
\Psi = \sum_{l=1}^{N}\sum_{m=-l}^{l} [a_{lm}r^{l} + b_{lm}r^{-(l+1)}]P_{lm}(cos\ \theta)e^{im\phi}
\label{eq:potential_spherical}
\end{equation} 
where the $a_{lm}$ and  $b_{lm}$ coefficients are determined by applying boundary conditions. The lower boundary is provided by the surface field values, and the upper boundary is given by the location of the maximum extent of the closed field corona, (known as the {\it source surface}). We denote this by  $r _{ss}$ and set its value to 3.5 R$_{\star}$. Beyond this radius, the field is purely radial and so $B_{\theta}(r_{ss}) = 0$ and $B_{\phi}(r_{ss}) = 0$. 

The three components of the magnetic field are then given by
\begin{equation}
B_r,B_{\theta},B_{\phi} = \left(-\frac{\partial}{\partial r} , -\frac{1}{r}\frac{\partial}{\partial \theta} , - \frac{1}{rsin(\theta)}\frac{\partial}{\partial \phi} \right) \Psi
\label{eq:mag_field}
\end{equation}

with solutions;
\begin{equation}
B_r = -\sum_{l=1}^{N}\sum_{m=-l}^{l} [la_{lm}r^{l-1} - (l+1)b_{lm}r^{-(l+2)}]P_{lm}(cos\ \theta)e^{im\phi}
\end{equation}

\begin{equation}
B_{\theta} = - \sum_{l=1}^{N}\sum_{m=-l}^{l} [a_{lm}r^{l-1} + b_{lm}r^{-(l+2)}]\frac{d}{d\theta}P_{lm}(cos\ \theta)e^{im\phi}
\end{equation}

\begin{equation}
B_{\phi} = - \sum_{l=1}^{N}\sum_{m=-l}^{l} [a_{lm}r^{l-1} + b_{lm}r^{-(l+2)}]\frac{P_{lm}(cos\ \theta)}{sin\ \theta}ime^{im\phi}
\end{equation}
where $P_{lm}$ are the Legendre polynomials. 

This field structure determines the distribution of open and closed field lines which may carry wind and prominences, respectively.

\subsection{Locations and Masses of Prominences}

Prominences may be supported at locations in the coronal magnetic field that correspond to stable mechanical equilibria. Stable points in the closed coronal field occur where there is a minimum in the gravitational potential following the direction of the field. These points require that
\vspace{2mm}
\begin{equation}
(\underline{B}\cdot\underline{\nabla})(\underline{g}_{eff}\cdot\underline{B}) < 0
\end{equation}
where $\underline{g}_{eff}$ is the effective gravity \citep{2001MNRAS.324..201J}.
Observed slingshot prominences have a temperature between 8000 and 9000 K \citep{1990MNRAS.247..415C}. Therefore, any closed field line which passes through a stable point is set to a temperature of $8500$ K. The maximum plasma density that can be magnetically supported in any stable point is given by \begin{equation}
    \rho_{max} = \frac{B^2}{\mu R_c |\underline{g}_{eff}| }
    \label{rho_max}
\end{equation}
where $R_c$ is the local radius of curvature \citep{2018MNRAS.475L..25V}. The prominence mass can therefore be found from this.

\subsection{Prominence Mass and Angular Momentum Loss}

Prominences are formed by mass flowing upwards from the stellar surface. If the loop that supports a prominence has a footpoint area $A_{foot}$, we can determine the rate at which mass flows into the prominence solely from the surface mass flux $ \rho_{\star} u_{\star}$ (where $\rho_{\star}$ and $u_{\star}$ are the plasma density and wind speed at the stellar surface). The mass flow rate through the footprint of a prominence is therefore
\vspace{2mm}
\begin{equation}
\Dot{M} =  \rho_{\star} u_{\star} A_{foot}.
\label{eq:8}
\end{equation}
The plasma density at the stellar surface is determined from the coronal temperature, 8.57 MK, and the base plasma pressure $p_0$ which we specify to be proportional to the local magnetic pressure $p_0 = \kappa_w B_0^2$, where the constant of proportionality $\kappa_w = 10^{-2.6}$ is chosen to reproduce the observed mass flux of the present-day solar wind \citep{2008ASPC..384..317C,10.1093/mnrasl/slw206}. 

As described in \citet{2019MNRAS.482.2853J}, the prominence mass will increase until it reaches the maximum value that can be supported by the magnetic field. The coronae of active stars are typically hot enough that this up-flow will be supersonic by the time it reaches the co-rotation radius where prominences are supported and as a result, information cannot travel back to the surface to shut off this flow. The prominence will be ejected and then reform in a limit-cycle. Thus the prominences act as temporary storage regions for the stellar wind, intermittently adding mass to the background wind when they are ejected. We can therefore identify the mass outflow rate in prominences with the rate $\Dot{M}$ at which mass flows into the prominence from the surface.
The prominence lifetime, $\tau_{prom}$, is then determined from the prominence mass as follows:
\begin{equation}
\tau_{prom} =  M_{max,prom} / \Dot{M}.
\label{eq:9}
\end{equation}

We note that in calculating the overall rate at which prominence mass is lost to the star, only the prominences that lie above the co-rotation radius will contribute. Any prominences below co-rotation will fall back to the star once they destabilise. We make a simple estimate of the minimum angular momentum loss rate from prominences by determining the angular momentum removed by the prominence mass from the site where it becomes destabilised. This neglects the extra torques imposed by the magnetic field as the prominence material moves outwards. Thus
\begin{equation}
\Dot{J} = \Dot{M} \omega_{\star} \overline{s}_{prom}
\label{eq:10}
\end{equation}
where $\overline{s}_{prom}$ is the cylindrical radius at which the prominence is supported and $\omega_{\star}$ is the stellar rotation rate.

\subsection{Wind Model}

In order to place these estimates of the mass and angular momentum losses from prominences in a broader context, we estimate also the contribution from the wind, using the WSA method \citep{1990ApJ...355..726W,2000JGR...10510465A} to find the stellar wind speed at a distance of 1 AU. The WSA wind speed depends on the expansion factor of the magnetic field between the stellar surface, $r=r_{\star}$, and the source surface, $r=r_{ss}$. Beyond the source surface, the field is assumed to expand radially. The expansion factor for any field line, $i$, is given by the relation
\begin{equation}
f_i = \frac{r_{\star}^2}{r_{ss}^2} \frac{B_i(r_{\star})}{B_i(r_{ss})}
\end{equation}
and the wind velocity for each field line at the source surface is given by
\begin{equation}
u_i[kms^{-1}] = 267.5 + \frac{410.0}{f_i^{2/5}}.
\end{equation}
This is set equal to the velocity at 1 AU.
The Parker wind solution is used to find the wind speed for the entire field line by finding the thermal velocity that matches the WSA velocity at 1 AU. Thus the temperature for the field line is found.

In a similar manner to the solution for the prominence upflow, we determine the plasma pressure at the base of the field line to be $p_0 = \kappa_w B_0^2$. The base temperature and pressure are then used to calculate the density at the base of the field line. From this, and the conservation of mass and magnetic flux through a flux tube, the mass loss rate from an open field line is $\dot{M}_i = \rho_i u_i A_i$, where $A_i$ is the area of a flux tube, $\rho_i$ is the density, and $u_i$ is the wind velocity. Integrating over a spherical surface the wind mass loss rate is
\begin{equation}
\dot{M} = \oint_{S_E} \rho_i u_i dS_i\ 
\label{eq:13}
\end{equation}
where $dS_i$ is the cross-sectional area of a flux tube.

The stellar wind carries away angular momentum from the Alfv\'en radius, where the wind velocity is equal to the Alfv\'en velocity,  $
u_A(r) = B(r) / \sqrt{\mu \rho (r)} $.
The total wind angular momentum loss is found by integrating over the Alfv\'en surface, $S_A$;
\begin{equation}
\dot{J} = \oint_{S_A} = \rho (\underline{u} \cdot \underline{n}) \Omega \overline{s}^2 dS_A
\label{eq:wind_ang_mom}
\end{equation}
where $\underline{n}$ is the normal in the outwards direction, $\overline{s}$ the Alfv\'en radius, and $\Omega$ is the stellar angular velocity.

\section{Prominences}
\label{sec:proms_main}

 \begin{figure}
    \includegraphics[width=\columnwidth]{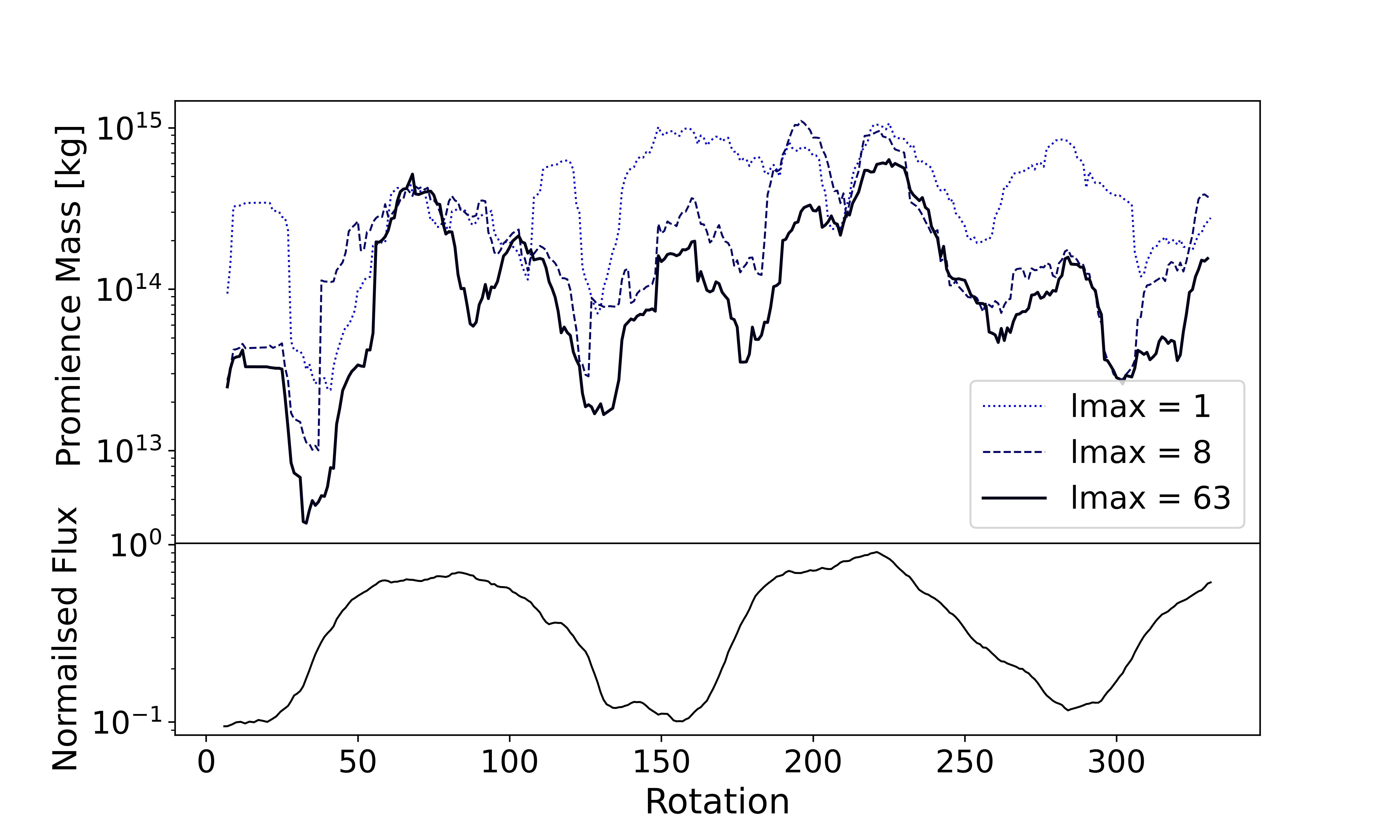}
    \caption{Top panel: predicted prominence mass at different resolutions averaged over 13 rotation periods. From highest to lowest resolution; the solid black line is solar resolution (LM63); the dashed black line is stellar resolution (LM08); the blue dotted line is dipole only (LM01). Bottom panel: normalised surface magnetic flux.}
    \label{fig:prom_mass}
\end{figure}

\subsection{Supported Mass}
\label{sec:supported_mass}
There are two factors that determine the total prominence mass that can be supported: the strength of the magnetic field and the availability of stable points. Both of these factors vary through the solar cycle. In order to isolate the effect of changing field strength, we plot the normalised magnetic flux in the lower panel of each plot. This also serves to highlight the periods of maximum and minimum of the cycle. An illustration of the geometry of the field and the location of stable points is shown in Figure \ref{fig:prom_models}. Since the stable points cluster at or beyond the co-rotation radius, which in this case is at about 2.7$R_\star$, the prominences are trapped in the largest-scale fields. Of all the multipole components, the dipole decays the most slowly with height and therefore at first glance it might be expected to dominate the prominence support. At cycle maximum, however, the lower-order terms, especially the quadrupolar term, can dominate \citep{2012ApJ...757...96D}.

The predicted prominence mass supported over the course of the magnetic cycle is shown in Figure \ref{fig:prom_mass}, with lines representing resolutions LM01, LM08, and LM63. Cyclic behaviour is difficult to identify in LM01, when only the dipole field is used, but is apparent in all higher resolutions, where the prominence masses show peaks at cycle maximum, with a secondary, lower-amplitude  peak at cycle minimum. 

At cycle maximum, the dipole term is weakest, but the alignment of the dipole axis with the rotation axis provides the greatest availability of stable points, and hence the maximum mass supported. At cycle minimum, the dipole is strongest, so although fewer stable points exist, they each support more mass, providing the secondary weaker maximum. It is at cycle minimum that the effect of the small-scale field is most apparent in reducing the prominence mass that can be supported. This is because a greater fraction of the magnetic flux closes at low heights where prominences cannot be supported. At cycle maximum this effect is reduced by the sheer number of stable points.

Over the course of a cycle, there is a large range in predicted mass values at all resolutions; just under two orders of magnitude for LM01 and over two orders of magnitude for LM63. At solar resolution, the range of {\it total} supported masses is approximately $10^{13} - 10^{15}$kg, which compares well with the masses of {\it individual} prominences observed on AB Doradus (2-6 $\times 10^{14}$kg) \citep{1990MNRAS.247..415C}.

\begin{figure*}
    \centering
    
        \includegraphics[width=8cm]{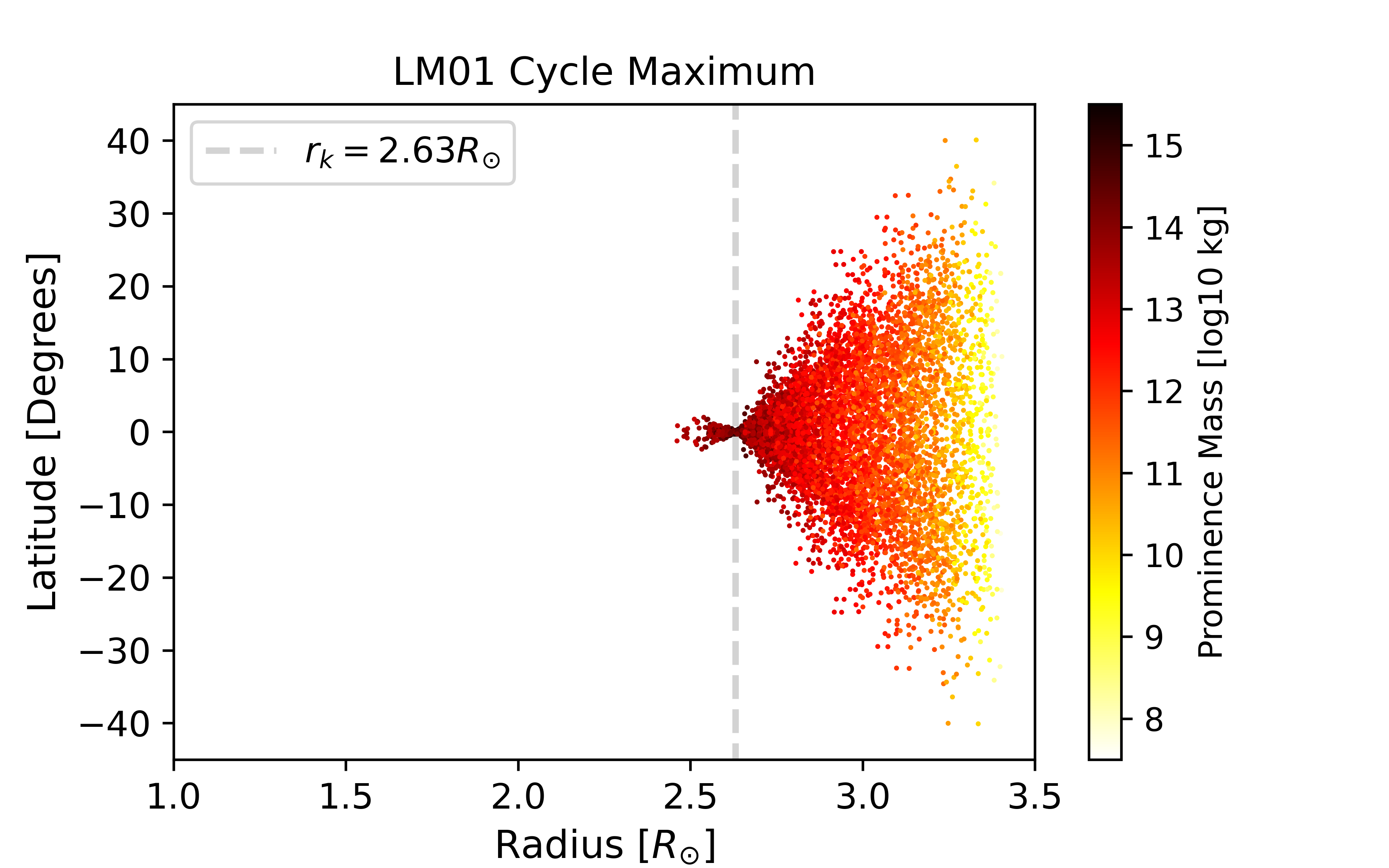}
        \includegraphics[width=8cm]{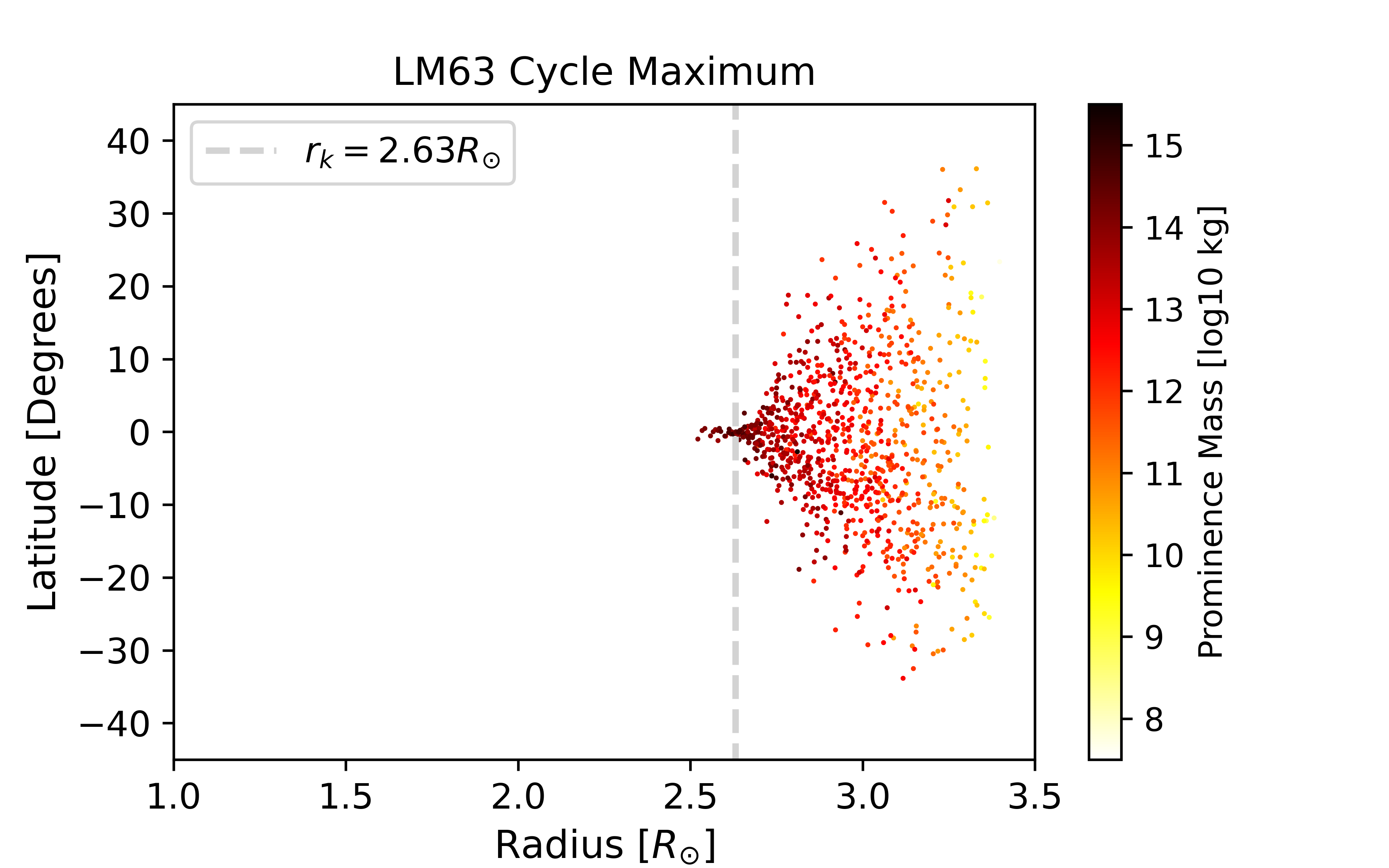}

    
        \includegraphics[width=8cm]{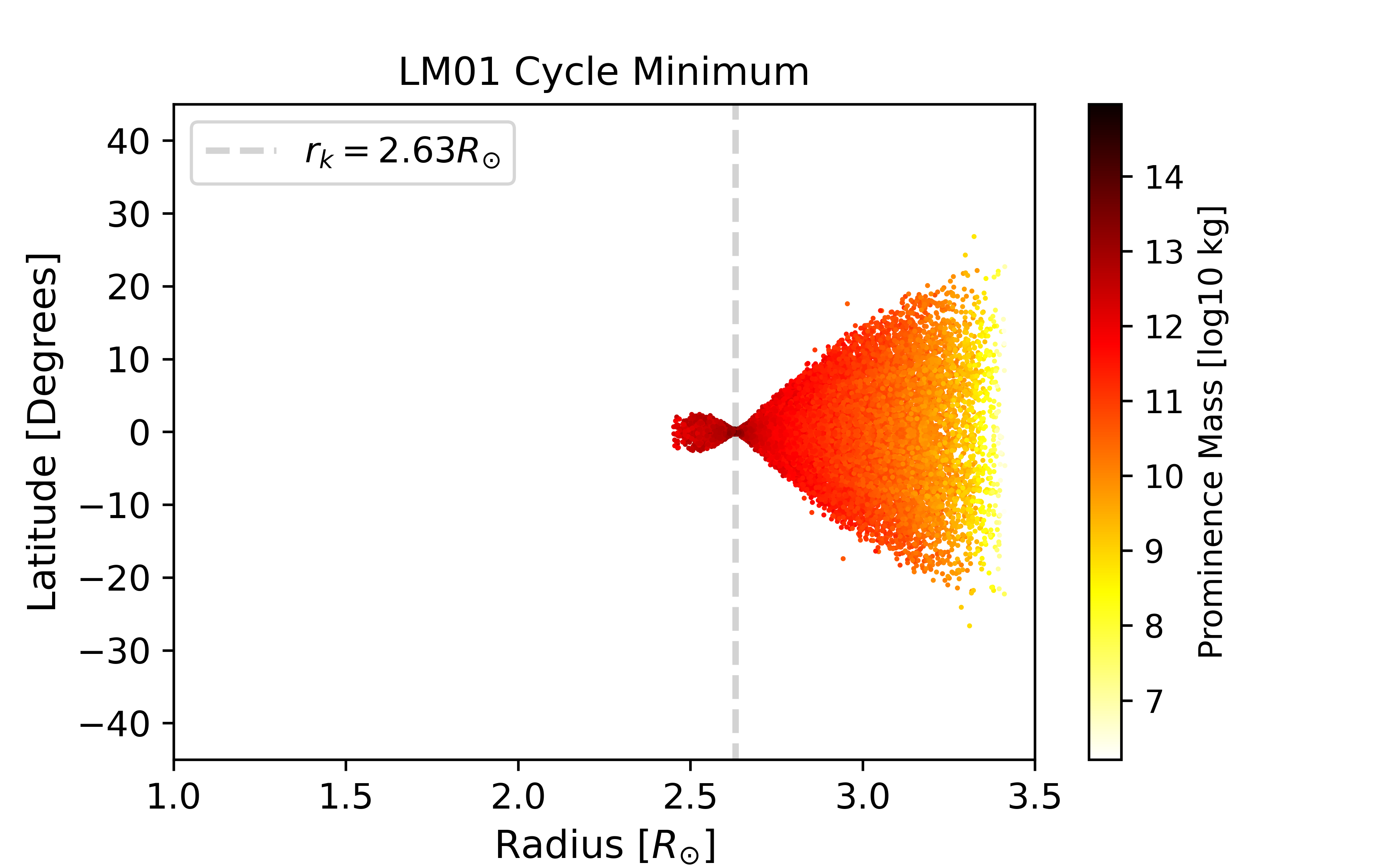}
        \includegraphics[width=8cm]{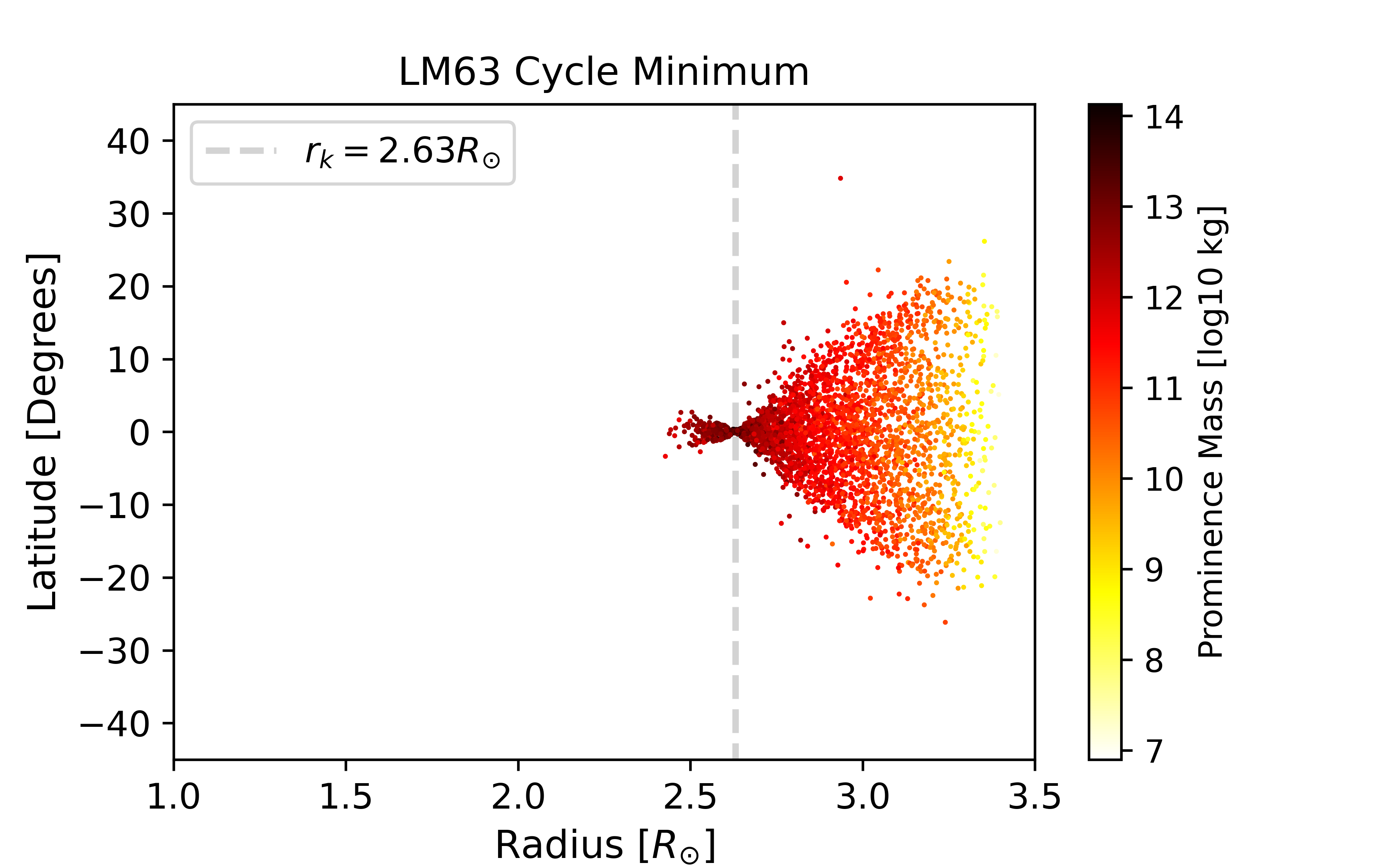}
    \caption{Distribution of slingshot prominence mass by latitude and radius at cycle maximum. Left plot is the prominence distribution for a dipole. Right plot is the prominence distribution at solar resolution. The dotted line shows the co-rotation radius. The colour denotes the amount of mass supported.}
    \label{fig:prom_lats}
\end{figure*}

\subsection{Latitudes}
\label{sec:latitudes}
In order to illustrate the interplay between the availability of stable points and the mass that can be supported in each, we show the distribution of prominences around the star as a function of latitude and radius in Figure \ref{fig:prom_lats}. Each plot shows prominence locations and masses for a range of surface maps around cycle maximum or cycle minimum, leaving out the majority of the rising and declining phases of the cycle. The plots on the top row are the prominences at cycle maximum, and the plots on the bottom row show prominences at cycle minimum. On the left are the predictions for LM01. The two plots on the right are predictions for the full resolution, LM63, maps. The colour bars represent mass, with the darker colours indicating higher mass. The most massive prominences are concentrated at the co-rotation radius.

By comparing the plots for LM01 and LM63 at either cycle maximum or minimum, it can be seen that the introduction of small scale field has little effect on the latitude and radius range of prominence sites, although there are fewer prominence-bearing stable points for LM63. This indicates that although the addition of small-scale field increases the field strength at the stellar surface, it reduces the number of prominences supported, and hence the total mass, as can be seen in Figure \ref{fig:prom_mass}. 
For both LM01 and LM63, the range of latitudes is greater at cycle maximum than at cycle minimum.  This suggests that the determining factor for prominence latitudes is the tilt of the dipole term relative to the rotation axis.

The most massive prominences lie at the co-rotation radius, where the effective gravitational term tends towards zero. This is also where the range of latitudes is smallest as prominences are confined to the equatorial plane. No prominences form beyond the source surface as there are no closed field lines beyond this point.

\subsection{Mass Loss}
\label{sec:mass loss}

 \begin{figure}
    \includegraphics[width=\columnwidth]{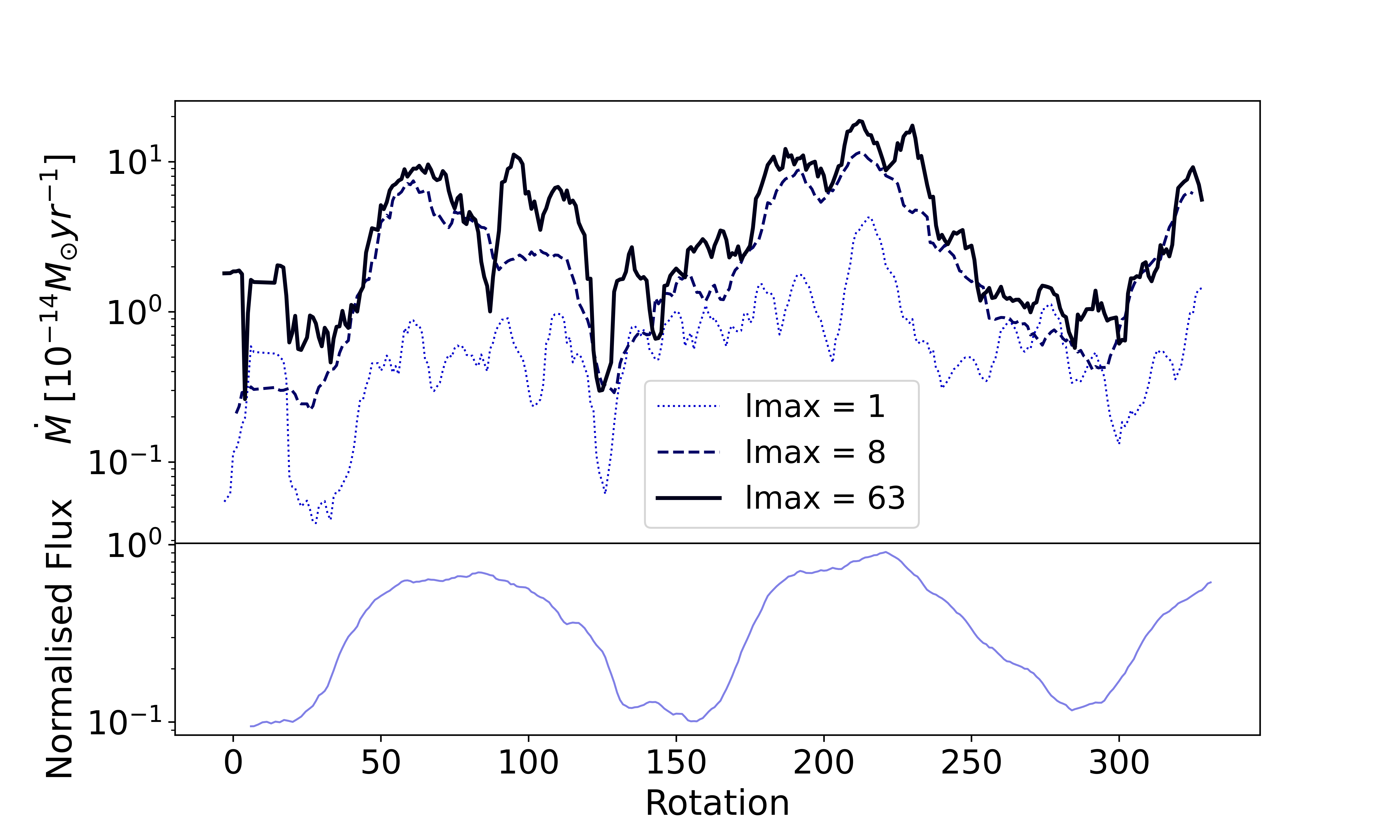}
    
    \caption{Top panel: predicted prominence mass loss rates at different resolutions. In order of increasing resolution the thick black line (LM01), thin black line (LM02), blue dashed line (LM08), blue dotted line (LM63). Bottom panel: normalised magnetic flux.}
    \label{fig:prom_mdot}

    \includegraphics[width=\columnwidth]{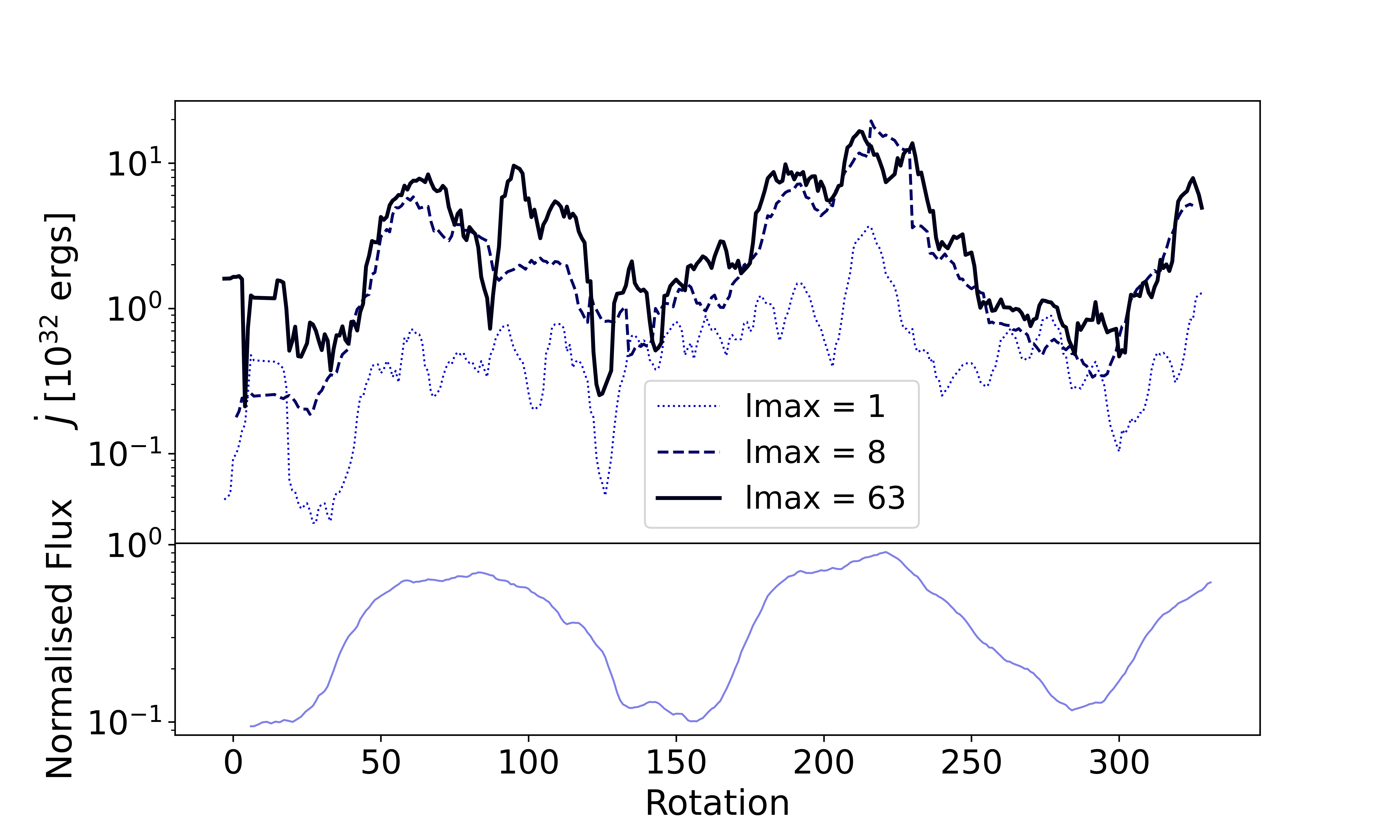}
    
    \caption{Top panel: predicted prominence angular momentum loss rates at different resolutions, averaged over a period of 13 rotations. In order of decreasing resolution; the thick black line is LM63; the dashed line is LM08; the blue dotted line is LM01. Bottom panel; normalised magnetic flux.}
    \label{fig:prom_jdot}
\end{figure}

Figure \ref{fig:prom_mdot} shows the predicted prominence mass loss rates. Including small-scale field results in an increase in the predicted mass lost through prominence ejection. This increase for LM08 and LM63 is most apparent near cycle maximum. LM01 prominence mass loss rates do not have strong variation associated with the magnetic cycle. There are short period oscillations that are most evident for LM01 which are possibly the result of Quasi Biennial Oscillations (QBOs) affecting the solar magnetograms. QBOs are oscillations in various measures of the the sun's activity, including sunspot number and neutrino flux, with periods of 0.6 - 4.0 years. Such oscillations and their origins have been a subject of study for many decades, but no consensus as to why they occur has yet been reached. For a comprehensive review of solar QBOs see \citet{2014SSRv..186..359B} and references therein.

Mass loss rates for LM08 and LM63 peak at cycle maximum. Unlike the behaviour of the supported prominence mass, there are no strong secondary peaks at cycle minimum. 
As LM01 has little to no cyclic variation, this suggests that the higher order field terms play a part in shaping the cycle. The increase of the surface field strength from higher order fields increases the plasma density at the prominence footprints, resulting in increased mass loss rates. 

There is a large range of mass loss rates, just under two orders of magnitude at all resolutions. At the highest resolution, LM63, the maximum and minimum mass loss rates are $2 \times 10^{-15} - 1 \times 10^{-13} $M$_{\odot} $yr$^{-1}$. The range of prominence mass loss rates for LM08 is very similar to that of LM63. Therefore, despite the loss of information due to low resolution, stellar resolution magnetograms are capable of reproducing the same range of prominence mass loss rates as solar resolution magnetograms.

\subsection{Angular Momentum Loss}
\label{sec:angular momentum loss}
The angular momentum loss rates are a product of the mass loss rates and the distance from the star where the prominence is ejected. This distance is usually around the co-rotation radius.
Shown in Figure \ref{fig:prom_jdot}, the maximum angular momentum loss occurs near cycle maximum for resolutions LM08 and LM63. As with the mass loss rates, LM01 does not show a strong cyclic variation, though the same short period oscillations are present.
The predicted angular momentum lost via prominences is lowest in LM01, just as with the predicted mass loss. LM08 and LM63 are similar in amplitude, though LM63 shows more variation.
As with the mass loss rates, angular momentum loss increases with the inclusion of small-scale field. There is a large difference between LM01 and the higher resolutions, but little difference between LM08 and LM63. This suggests that predicted prominence angular momentum loss rates are recovered well with stellar resolution magnetograms, while the dipole term provides a lower limit.

As the heights of prominences do not change significantly over the course of a cycle, the peaks in angular momentum loss are due to the increase in mass loss, as evidenced by their similar shape.
The range of angular momentum loss for LM63 is $0.1 - 6 \times 10^{32}$ erg, which is slightly larger than the range for LM08. The range of values for the predicted angular momentum loss via prominences is more than an order of magnitude. This very large scatter is a result of the stellar cycle, and much larger than the difference between typical solar and stellar resolutions.

\section{Wind}
\label{sec:wind main}
\subsection{Wind Mass Loss}
\label{sec:wind mass loss}
\begin{figure}
    \includegraphics[width=\columnwidth]{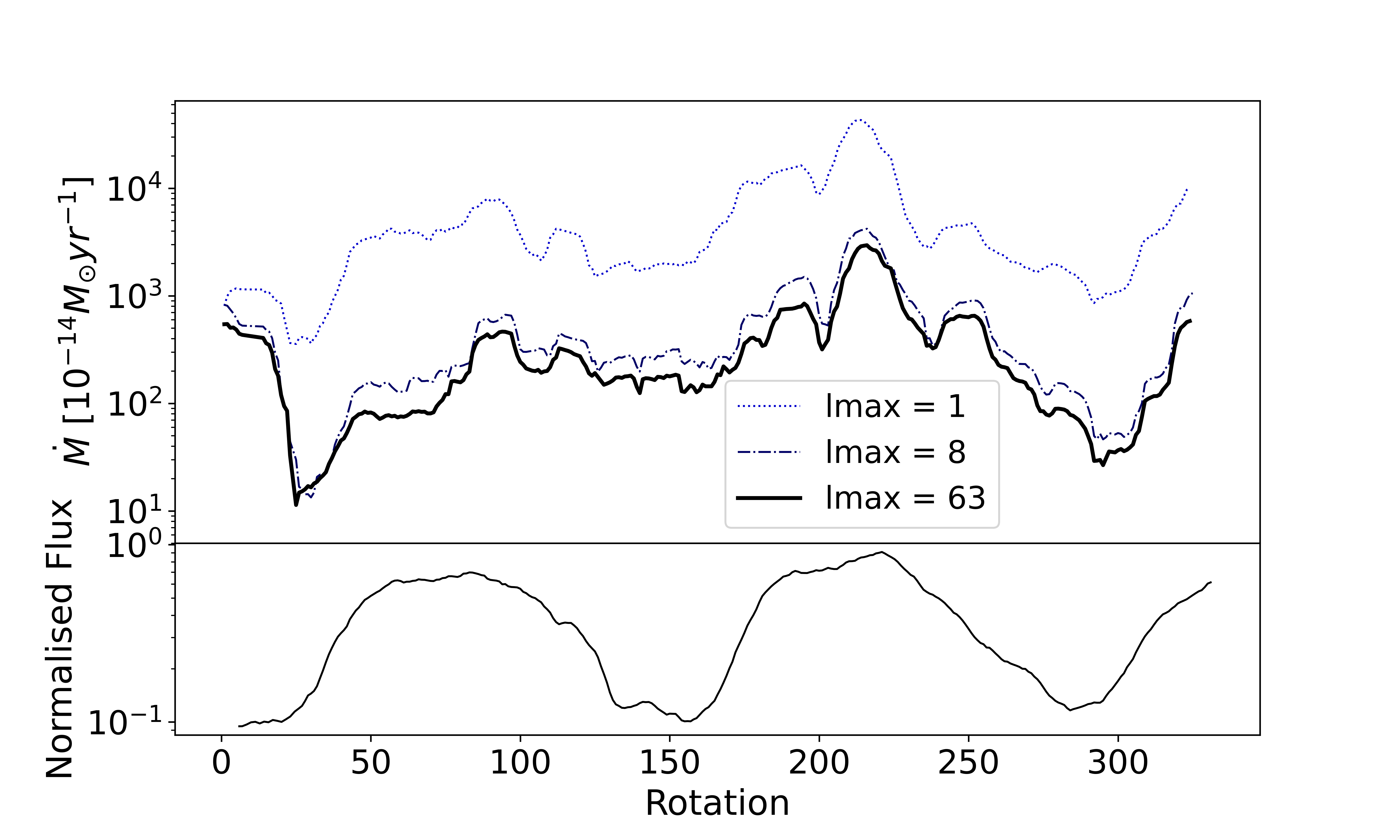}
    \caption{Predicted wind mass loss rates at different resolutions. in order of decreasing resolution, the thick black line is solar resolution (LM63); the dashed black line is stellar resolution (LM08); the dotted blue line is dipole field (LM01). The panel underneath is normalised magnetic flux.}
    \label{fig:wind_mdot}
    \includegraphics[width=\columnwidth]{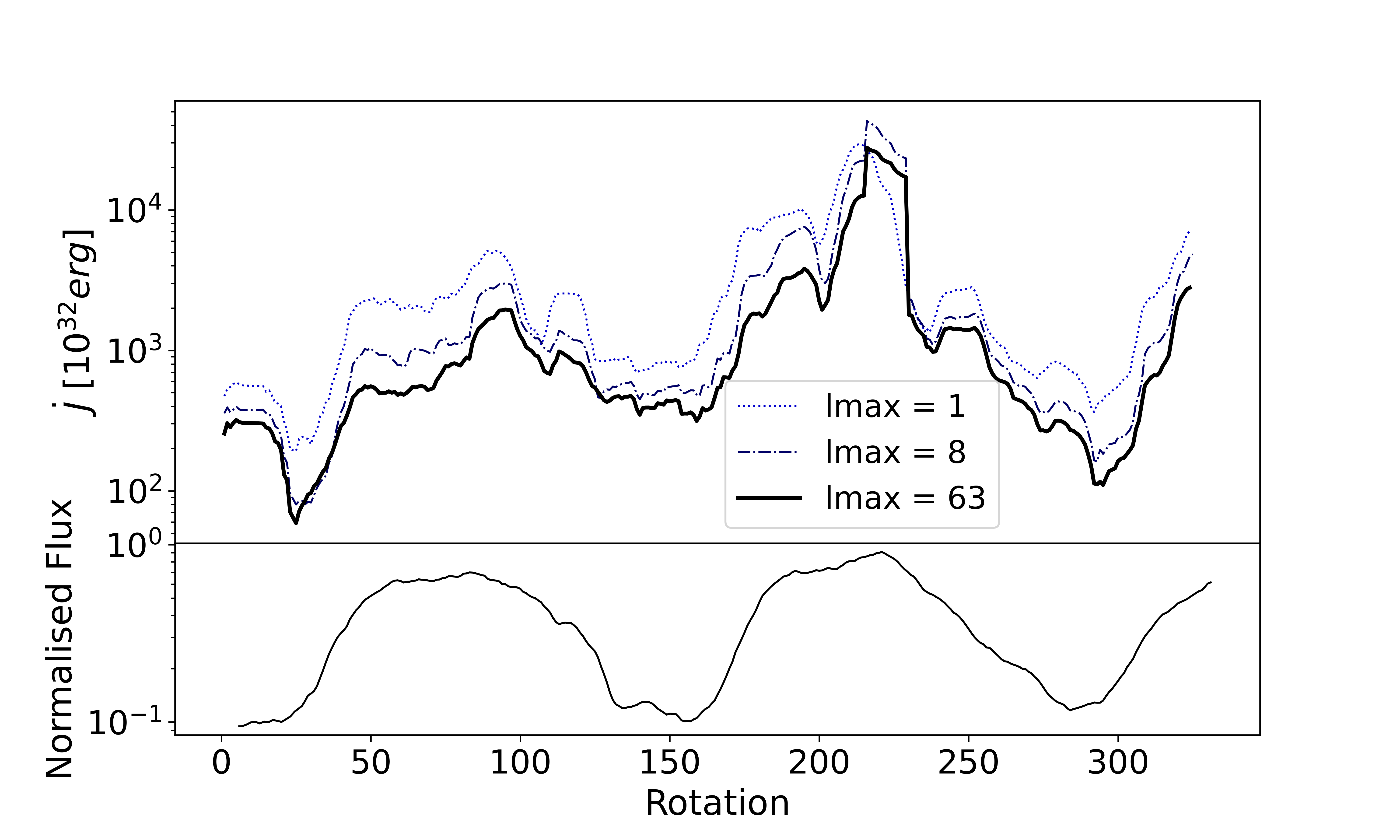}
    \caption{
    Predicted angular momentum loss rates for the wind.
    For each plot: in order of decreasing resolution, the thick black line is solar resolution (LM63); the dashed black line is stellar resolution (LM08); the dotted blue line is dipole field (LM01). The panel underneath is normalised magnetic flux.}
    \label{fig:wind_jdot}
\end{figure}

The cyclic variation in wind mass loss rates at various resolutions is shown in Figure \ref{fig:wind_mdot}. Unlike the prominence mass loss rates discussed in section \ref{sec:mass loss}, the shapes of the variations at all resolutions are virtually identical, though of different magnitudes, with peaks at cycle maximum. The LM01 mass loss rates are an order of magnitude greater than the LM08 and LM63 mass loss rates, whereas the latter two are closely matched. This is consistent with the findings of \citet{2017MNRAS.465L..25J} who found that using only the dipole component of solar magnetograms gave an overestimate of wind mass loss rates. The surface field strength of the star increases at higher resolution due to the inclusion of small-scale flux. This inclusion results in a higher base density at the surface of the star and increased mass flow along field lines.



The wind mass loss rates are greater than those from prominences. Over the course of the cycle, the range of wind mass loss values for LM63 are an order of magnitude, which is much greater than the difference between LM08 and LM63, indicating that stellar resolutions are reasonable for reconstructing higher resolution wind mass loss rates.

\subsection{Wind Angular Momentum Loss}

The cyclic variation in angular momentum loss rates from the wind is shown in Figure \ref{fig:wind_jdot}. As in section \ref{sec:wind mass loss}, the shape of the cyclic variation is similar at all resolutions, though there is not a large difference in values between LM01 and LM08. Similarly to the wind mass loss rates, the effect of increasing the magnetogram resolution is to reduce the angular momentum lost in the stellar wind.

Wind angular momentum loss rates, given by equation \ref{eq:wind_ang_mom}, are a product of the mass loss and the Alfv\'en radius. The variation in the mass loss rates are dependent on the base density at the stellar surface, which is turn depends on $B^2$. As $B^2$ increases, the Alfv\'en radius decreases, yet we still see a very large variation in the angular momentum loss rates.

The wind angular momentum loss rates are similar at all resolutions, much more so than the mass loss rates. This is because the Alfv\'en radius increases in such a way as to compensate for the wind mass loss. Therefore, the result is that the variation in angular momentum loss within a cycle is much greater than the difference between the LM01 and LM63 cycles.

\section{Discussion \& Conclusions}
Observations of stellar prominences and indeed stellar winds are relatively few in number and sparse in time. Repeated observations show variation for individual stars that are comparable to the difference in observations for different stars. This paper aims to examine two sources of variability to determine whether the variation could be intrinsic, caused by magnetic cycles or the presence of undetected small-scale field.
From the figures in section \ref{sec:proms_main} and section \ref{sec:wind main}, the greatest range in values for both mass and angular momentum loss are due to cyclic variations.
The relative contributions of magnetogram resolution and cyclic variability are discussed below.

\subsection{The influence of the small-scale field}
The resolution of a magnetogram determines how much small scale flux is resolved. At the LM01 resolution the dipole term is the only one seen, meaning that features such as spots and active regions are overlooked. However, the dipole is an important feature in following the progression of the magnetic cycle. In the solar case, the dipole term will be aligned with the solar rotation axis during cycle minimum and will drift closer to the equator near cycle maximum due to the influence of sunspots. At this point the dipole field goes through a reversal before again becoming aligned with the rotation axis, though in opposite polarity. 

The surface field shapes the coronal magnetic field. When aligned with the rotation axis, a sufficiently strong dipole field on a fast-rotating star could theoretically support a torus of prominences in the equatorial plane, if the co-rotation radius lies within the source surface radius.
As the dipole field becomes more inclined away from the rotation axis, the torus morphs to become two discrete bands either side of the star. A highly inclined dipole is able to support these prominence bands to higher latitudes than an aligned dipole, as shown in the left side of Figure \ref{fig:prom_lats}.
This finding is supported by \citet{articlewaugh}, who predicted the prominence formation sites for a number of M-dwarfs from the corresponding ZDI maps.

The locations at which prominences can be supported in the corona change, however, as more small-scale field is included. At cycle minimum when the dipole term is aligned with the rotation axis, there is no longer the possibility of a perfect torus of prominences around the equator due to the influence of higher order fields, but the prominences that do form are bound to similar latitudes as in the aligned dipole case. At cycle maximum, however, prominences are again able to form at higher latitudes. As shown in Figure \ref{fig:prom_lats}, the range of prominence latitudes at cycle minimum and maximum are the same for resolutions LM01 and LM63. The greatest change caused by the addition of small scale field is the reduction in the number of prominences that form.

The locations of prominences form a funnel shaped distribution in latitude and radius. They are highly concentrated at the co-rotation radius before spreading out until the source surface radius is reached. The colour bars in Figure \ref{fig:prom_lats} show the amount of mass supported at each prominence site. The most massive prominences are situated at the co-rotation radius where the effective gravity tends towards zero.

Although the inclusion of small scale surface flux limits the number of stable points in the coronal field, the increased field strength provides more support for the remaining prominences. This additional support means that each prominence is capable of containing more mass. A star with a solar-like cycle will generate small scale flux leading up to, and peaking, at cycle maximum. This is the reason why at cycle maximum, the prominence masses are similar at all resolutions.

Contrary to the supported prominence mass, the mass loss rates and angular momentum loss rates for prominences increase with the addition of small scale field, although this increase is largely confined to the very lowest-order modes. LM08 and LM63 are very similar in that LM63 follows the shape of LM08 in Figure \ref{fig:prom_mass} and Figure \ref{fig:prom_lats}, whereas LM01 is consistently lower. LM01 shows very little of the variation expected with a magnetic cycle as it does not have clearly defined maxima and minima as in LM08 and LM63. From this it can be concluded that the dipole term is insufficient to predict prominence mass and angular momentum loss rates on its own.

Consistent with the results of \citet{2017MNRAS.465L..25J} we find a modest decrease in wind mass and angular momentum loss rates with the inclusion of small-scale field, but this is limited to the very low-order modes. 

The influence of small scale field on the wind angular momentum loss is less than that on the wind mass loss rates. This is because including small scale field increases the Alfv\'en radius, which compensates somewhat for the decrease in mass loss rates. 
In this model, we have used solar magnetograms with amplified field strength. If the number of active regions were increased in line with a very active you star, we would expect there to be fewer prominences. This follows from the argument that the inclusion of small-scale flux reduces the number of stable points in the corona out at the co-rotation radius. Increasing the number of active regions increases the amount of small-scale flux. With enough active regions present, it would be increasingly unlikely that slingshot prominences could form. Therefore, despite the increase in surface field strength allowing for larger prominences, there may be an activity level beyond which it would be almost impossible for a star to host slingshot prominences, and the total mass of such structures would tend towards zero. However, such a star could still form prominences closer to the surface of the star, although these types of prominences are not considered in this work.

\subsection{The effect of magnetic cycles}

At dipole resolution, LM01, the mass supported in prominences varies by an order of magnitude within the cycle, with peaks at both cycle maximum and minimum. At the highest resolution, LM63, the mass varies by two orders of magnitude, and the peaks at cycle maximum are larger than those at cycle minimum. 
This double-peaked behaviour at all resolutions is unexpected, but likely a simple consequence of field geometry. The dipole field has two components; the axial and the equatorial harmonics. At cycle minimum, the magnitude of the axial mode is at its greatest, as the dipole is aligned with the stellar rotation axis \citep{2012ApJ...757...96D}. This is the field which can support a torus of prominences around a star. At cycle maximum however, it is the equatorial dipole mode which is at its greatest magnitude. In this configuration, the equatorial to axial dipole ratio is large enough to provide the prominence support necessary to show this secondary peak in the prominence masses. Such behaviour is most prominent with the dipole term, but persists to higher resolutions. This behaviour was identified, but not explicitly discussed, by \citet{2020MNRAS.491.4076J} in examining the behaviour of prominences in a dipole field. Figure 3 in \citet{2020MNRAS.491.4076J} shows that, for a constant strength dipole, the minimum mass supported in prominences occurs when the latitude of the dipole axis is around 20$^{\circ}$ from the equator. The largest mass supported occurs when the dipole axis is aligned with the rotation axis, and a secondary peak occurs when the dipole is fully equatorial.

The prominence mass and angular momentum loss rates do not show this secondary peak. In both Figure \ref{fig:prom_mdot} and Figure \ref{fig:prom_jdot}, LM01 shows no cyclic behaviour, but LM08 and LM63 are very similar in values and show clear cyclic variation.
The maximum prominence mass loss rates occur at cycle maximum, because of the increased surface field strength. By equation \ref{eq:8}, prominence mass loss rates are proportional to the density at the surface of the star, which is related to the square of the magnetic field strength. The surface field strength increases with the emergence of spots and active regions associated with cycle maximum. Therefore, the mass flow rate into the prominences increases towards cycle maximum, thereby increasing the rate at which prominences are filled, destabilised, and ejected. The increased field strength also allows more mass to be supported in the prominences during cycle maximum.
In this paper, only prominence material that lies above the co-rotation radius is considered when determining the mass and angular momentum loss rates. Any prominences that form below this radius would fall back to the star once they destabilise.
The prominence angular momentum loss rates also peak at cycle maximum. The variation over the course of the cycle is similar to the mass loss rates. This is expected as the angular momentum loss rates depend on the mass loss rates. The other variable to consider is the radius at which the prominences are ejected, but by Figure \ref{fig:prom_lats}, it can be seen that the range of radii over which prominences form are the same regardless of cycle or resolution.

The wind mass and angular momentum loss rates show cyclic behaviour at all resolutions. Between resolutions, the behaviour of each cycle is very similar, varying only in magnitude. This is a stark contrast to the behaviour of prominence masses at different resolutions.
The wind mass loss rates peak at cycle maximum. From equation \ref{eq:13} it is seen that the wind mass loss rates are dependent of the density at the base of the field line. As argued above, the cyclic variation in base field strength is responsible for some of the variation seen in the wind mass loss rates. From {\it in-situ} measurements of the solar wind, the amount of open flux, varies throughout the magnetic cycle, the greatest being at cycle maximum. 

On the Sun, the observed wind mass loss rates vary only by a factor of between 2 and 5 between cycle minimum and maximum, which is attributed to the small change in the amount of open flux over the solar cycle \citep{10.1111/j.1365-2966.2011.19428.x}. Since the model in this paper is based on solar data, albeit with an increased source surface radius and field strength, it may be surprising that there is a greater variation in wind mass loss rates for this young Sun model. However, the standard deviation around the mean is only a factor of 5.

There are several possible explanations to explain the difference. Firstly, the model has a larger source surface radius than the Sun which leads to a smaller percentage of open to total flux. Secondly, the surface field strength is increased by factor of approximately 160 across the star. This leads to an increase in base density which varies as $B^2$. Finally, solar wind mass loss rates use {\it in-situ} measurements which will be very different from models as they only sample one small part of the wind at any given time. In order for models to recreate the solar wind values, a variable source surface must be used, which is beyond the scope of this work \citep{2018ApJ...864..125F}.

The level of variation from both wind and prominence mass and angular momentum loss rates for this model are greater than any variation due to the inclusion or neglect of small-scale fields. Prominence mass and angular momentum loss rates fluctuate by more than an order of magnitude over the course of the magnetic cycle, while wind fluctuations can exceed two orders of magnitude. Herein lies an explanation for the wildly varying values determined from stellar observations. \citet{2020MNRAS.491.4076J} used ZDI maps of AB Dor to predict prominences that vary in mass from $0.3 - 29 \times 10^{14}$ kg, with prominence mass loss rates of $0.4-24 \times 10^{-14} $M$_{\odot}$yr$^{-1}$. The prominence mass values obtained in this work are slightly lower, and the range of mass loss rates slightly smaller than those predicted for AB Doradus. Some differences are expected, however, as this model uses the solar flux distribution to model a star with a shorter rotation period. The results here suggest that magnetic cycles can have as large an effect on prominence masses and distributions as variations in stellar mass or rotation rate.

\section*{Data Availability}
The magnetograms used in this work originate from the US National Solar Observatory, Kitt Peak. The archival data may be found in the NSO historical archive at \url{https://nispdata.nso.edu/ftp/kpvt/synoptic/mag/}. Supplementary data is available at \url{10.17630/1f1c12c0-fc29-4571-a11e-e0d45a05d713}

\section{Acknowledgements}

The authors acknowledge support from STFC consolidated grant number ST/R000824/1. 

\bibliographystyle{mnras}
\bibliography{PaperJan2022} 

\bsp	
\label{lastpage}
\end{document}